\documentclass[smallextended]{svjour3} 
\usepackage{tikz}
\usepackage{graphics,graphicx,epsfig}
\usepackage{amsmath}
\usepackage{amsfonts}
\usepackage{comment}
\def\identity{\leavevmode\hbox{\small1\kern-3.8pt\normalsize1}}

\newcommand{\be}{\begin{eqnarray}}
\newcommand{\ee}{\end{eqnarray}}

\newcommand{\Tr}{\mathrm{Tr}}

\renewcommand{\epsilon}{\varepsilon}

\usepackage{graphicx}
\usepackage{tikz}
\usetikzlibrary{arrows}
\usepackage{graphicx}
\begin{document}

\title{On the Security of Semi Device Independent QKD protocols}

\author{Anubhav Chaturvedi \and 
        Maharshi Ray \and 
        Ryszard Veynar \and
        Marcin Paw{\l}owski}

\institute{ A. Chaturvedi \at
Institute of  Theoretical Physics and Astrophysics, National Quantum
Information Centre, Faculty of Mathematics, Physics and Informatics, University of Gda\'nsk, Wita Stwosza 57, 80-308 Gda\'nsk, Poland \\
Center for Computational Natural Sciences and Bioinformatics, Center for Security, Theory and Algorithmic Research, International Institute of Information Technology, Gachibowli, 500032, Hyderabad, India.\\
\email{anubhav.chaturvedi@research.iiit.ac.in}
\and 
M. Ray \at 
Center for Computational Natural Sciences and Bioinformatics, Center for Security, Theory and Algorithmic Research, International Institute of Information Technology, Gachibowli, 500032, Hyderabad, India. 
\and 
R. Veynar \at
Institute of  Theoretical Physics and Astrophysics, National Quantum
Information Centre, Faculty of Mathematics, Physics and Informatics, University of Gda\'nsk, Wita Stwosza 57, 80-308 Gda\'nsk, Poland 
\and 
M. Paw{\l}owski \at
Institute of  Theoretical Physics and Astrophysics, National Quantum
Information Centre, Faculty of Mathematics, Physics and Informatics, University of Gda\'nsk, Wita Stwosza 57, 80-308 Gda\'nsk, Poland 
}

\maketitle

\begin{abstract}
While fully device-independent security in (BB84-like) prepare and measure Quantum Key Distribution (QKD) is impossible, it can be guaranteed against individual attacks in a semi device-independent (SDI) scenario, wherein no assumptions are made on the characteristics of the hardware used are made except for an upper bound on the the dimension of the communicated system. Studying security under such minimal assumptions is especially relevant in the context of the recent {\it quantum hacking} attacks wherein the eavesdroppers can not only construct the devices used by the communicating parties but are also able to remotely alter their behavior. In this work we study the security of a SDIQKD protocol based on the prepare and measure quantum implementation of a well-known cryptographic primitive, the Random Access Code (RAC). We consider imperfect detectors and establish the critical values of the security parameters (the observed success probability of the RAC and the detection efficiency) required for guaranteeing security against eavesdroppers with and without quantum memory. Furthermore we suggest a minimal characterization of the preparation device in order to lower the requirements for establishing a secure key.
\end{abstract}


\section{\label{sec:level1}Introduction}

In standard quantum key distribution (QKD) protocols the security proofs assume that the parties have access to the correct and exact specifications of the devices used therein. 
This assumption is rather problematic. First, the principle problem lies at the heart of quantum formalism. Generally, these devices are used for either state preparation or measurement. The quantum formalism provides mathematical abstractions for states and measurements but no direct way to infer about them individually. The only interface for any inference about the states and measurements is via the Born rule which combines states and measurements and yields the probability of outcomes, which is compared to experimental results. So a full characterization of the devices is sufficient to warrant security but it requires that each device and its components are individually tested several times to gather enough statistics in order to warrant trust. This is an extremely tedious task, instead we end up trusting the manufacturer of the devices, which may be not the best idea. For instance, the supplier can install back-doors, that enable him to compromise the security without being detected. Recently a lot of attention has been drawn to NSA which convinced RSA Security to set as a default in their products Dual\_EC\_DRBG pseudo-random number generator which is in-turn known to have such a back-door\cite{NSA}. Moreover, even if the manufacturer is honest, recent advances in {\it quantum hacking} \cite{lydersen2010hacking} show that the adversary can remotely influence the behavior of the devices during the protocol, effectively changing their characteristics thereby hampering the security of the protocol. To cope with this issue the device independent (DI) approach has been introduced wherein the key idea is that if the parties violate a Bell (or some Bell-like) inequality then, regardless of how their devices managed to do this, they can establish secure communication. Although the term ``DI" was first used in \cite{DI1} the idea can be tracked back all the way to the original Ekert's paper \cite{E91}. Unfortunately, completely DI QKD is extremely arduous to realize in practice and so far no experimental group has been able to do this. The main reason for this is so-called detection efficiency loophole \cite{Pearle}, which states that if the probability of registering a particle by the detectors used in the experiment is below a certain (usually very high) critical value then the results of the experiment are inconclusive, in other words: the possibility of a local-realistic description of its results cannot be ruled out. Ruling out the possibility these descriptions is a necessary, although not always sufficient, condition for DI security \footnote{Necessity stems from the fact that if the experiment could be described by classical model it must be insecure as classical key distribution is impossible without additional assumptions on computational power of the eavesdropper. Whether or not this condition is sufficient depends on the details of the protocol and the power given to the eavesdropper.}. Another problem faced in this scenario is that it can be applied only to protocols based on entanglement which are much more complicated than their prepare-and-measure counterparts like the BB84 \cite{bb84original}.

These two issues are addressed in the semi-device independent (SDI) approach \cite{pawlowski2011semi}. Here a prepare-and-measure scenario is considered and again no assumptions are made on the inner specifications and the working of the devices used. Prefix ``semi" is warranted by the fact that an upper bound on the dimension of the communicated system is assumed. Assuming this bound is well justified for both honest and dishonest manufacturer. In latter case the parties can study the devices delivered and, while it is almost impossible to fully characterize them, it is much easier to establish the effective dimension of the Hilbert space in which the states are being prepared. When the supplier is honest but the protocol is subject to a quantum hacking attack, the limitations on the technology available to the eavesdropper make the task of increasing the channel capacity extremely difficult. In fact, to our knowledge, all the quantum hacking attacks published so far did not increase this capacity. Also, because only one side employs the detectors, the requirements on their efficiency are lower than in the DI case. Another relaxation of the DI paradigm is measurement-device independent (MDI) scenario \cite{MDI}\cite{MDI2}\cite{Am2}, wherein three devices are used: two communicating parties, with perfectly characterized hardware are sending the particles to the third which makes the measurements. No assumptions are made on the characteristics of the third device. The difference between MDI and SDI scenarios is that the former one is more complicated (i.e. requires more devices and more sophisticated measurements) and does not allow for any changes in the preparation devices (which is a big disadvantage as even small changes can lead to the loss of security \cite{mAlice}). On the other hand, it was shown \cite{MDI}  that MDI scenario thwarts quantum hacking attacks for any efficiency of the detectors.

The aim of this paper is to establish the security condition in the SDI case i.e. finding the critical values of the security parameters required in order to establish a secure key. In the case of Bell inequalities, given an observed value of detection efficiency (greater than the critical detection efficiency), if the parties witness a Bell violation above a certain threshold, they are sure that the system they share must be non-local (or entangled if quantum theory is assumed). Similarly we find the threshold above which the SDIQKD protocol is secure based on the threshold value of some observed parameter (for a given certain value of observed detection efficiency). We start by defining the classes of attacks against which we want to be secure. We consider individual attacks in which eavesdropper may or may not have access to quantum memory. Then we take the most basic SDIQKD protocol based on $(2\to1)$ QRAC \cite{pawlowski2011semi} and find the security conditions required against such attacks. Next we propose a modification of this protocol (basing it on the $(3\to1)$ QRAC instead) which substantially reduces these security requirements. Furthermore, we suggest a minimal characterization of the preparation device that substantially lowers the requirements on the security parameters.
\section{Device controlling attacks}

In \cite{lydersen2010hacking,Makarow4}, the authors gave a simple description of the device controlling attacks based on the detection efficiency loophole and experimentally demonstrated the same. Assuming that Eve has perfect detectors while Bob's detectors have an average efficiency of $50\%$. Eve intercepts the signal sent from Alice to Bob as a part of the BB84 protocol and measures it. Following which Eve encodes her detection results into specially tailored bright pulse of light and resends it to Bob. Because of the physical properties of the signal and the physical implementation of Bob's detectors, Bob obtains an outcome only if he measures in the same basis as Eve. This implies that Bob's detectors work perfectly (with 100$\%$ efficiency) conditioned on the even that Eve's and Bob's settings are the same and not work at all otherwise. On an average Bob's detectors work with $50\%$ efficiency which does not raise any suspicions. After the raw key exchange, Bob and Eve have identical bit values and basis choices which after sifting, error correction and privacy amplification made via classical communication allows Eve to get the identical final key as Alice and Bob. In this way Eve, by active control of Bob's detectors can secretly learn the exchanged key. More about this type of control can be found in \cite{Makarov3}.

In \cite{mAlice} a different approach is presented. Here Eve apart from exploiting her possibility of interfering during the calibration of Alice's device introduces a slight modification in it.

These examples highlight the need for more general security conditions where Eve is assumed to be substantially powerful i.e. she can not only design all the devices used in the protocol but can also actively control them. However there are natural limits to what she can do. Her modifications must not be significant so as to avoid detection. Apart from this we assume that Eve cannot make the preparation device use additional degree of freedom of the communicated system to encode more information. Hence in the SDI scenario we assume that the eavesdropper can mold the characteristics of the devices used as well as actively control them during the protocol but cannot increase the dimension of the system sent by Alice.

\section{\label{sec:level1}SDIQKD protocol}
The DIQKD protocol bases its security on violation of a Bell inequality \cite{bell1964einstein} associated with the scenario. The key rate, in this case, is maximized by reaching the quantum bound of this inequality. On the other hand, SDIQKD is a prepare and measure key distribution protocol wherein the dimension of the communicated system is upper bounded \cite{pawlowski2011semi}. Specifically, this limitation is only for the dimension of the signal emitted by Alice's device and doesn't hold for what Bob's device is receiving. This minimal restriction is tremendously advantageous for Eve, which in-turn captures the fact that in device controlling attacks, the pulse sent by Eve to Bob's lab could carry substantially more information than just one bit.
The SDIQKD scheme bases its security on beating the classical bound on the winning probability (efficiency) of a related communication complexity task. In \cite{pawlowski2011semi} the task used was a $(2\to 1)$ Quantum Random Access Code (QRAC)\cite{qrac}, a prepare and measure quantum implementation of a well-known cryptographic primitive, the Random Access Code (RAC).
In a $(2\to 1)$ QRAC Alice encodes her two input bits $a_0,a_1\in\{0,1\}$ into a qubit $\rho_{a_0,a_1}$. Bob gets an input bit $b\in\{0,1\}$ and chooses his projective measurements $M_b^B$ based on it, where $B\in \{0,1\}$ is the outcome of his measurement. Bob's task is to return $B=a_b$ i.e. guessing the $b$th bit of Alice. The measure of success in this task is the probability with which Bob is able to guess correctly, we denote it by $P_B$.

\begin{figure}[htp]
\centering
\resizebox{.95\linewidth}{!}{\includegraphics[]{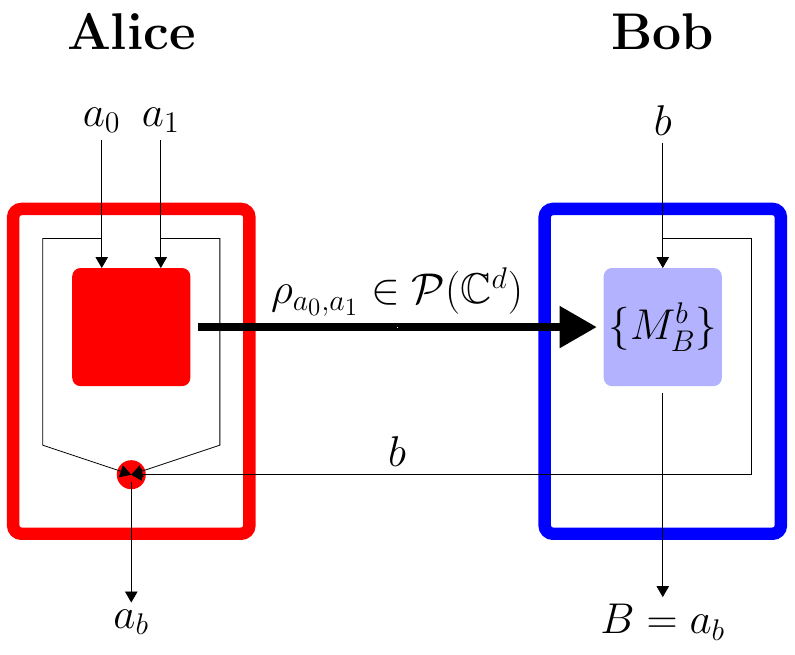}}
\caption{A single round of the SDIQKD protocol based on $(2\to 1)$ QRAC. Here the larger boxes represent Alice's and Bob's labs of which their preparation and measurement devices respectively (represented by smaller boxes) are a part of. The thinner lines represent classical communication channels and thick lines represent quantum communication channels which in this case has a bounded capacity. The SDIQKD protocol contains the $(2\to1)$ QRAC along with a classical communication channel carrying Bob's input $b$ and the classical post-processing at Alice's end (represented by a small disk) required to output $a_b$ using $a_0,a_1,b$ as inputs.}
\label{21}
\end{figure}

The SDIQKD protocol introduced in \cite{pawlowski2011semi} comprises of many repetitions of $(2\to 1)$ QRAC. In each of them $a_0,a_1$ and $b$ were chosen randomly by Alice and Bob, respectively. After this Bob announces his choice of $b$ for each round. The bit $a_b$ forms the bit of key for that round. Alice knows it, since it is one of the bits she herself randomly generated. Bob has some information about $a_b$. Probability for Bob to obtain (for fixed settings $a_0,a_1$ and $b$) the result $i$ is $P(B=i|a_0,a_1,b)=tr(M_{b}^{B=i}\rho_{a_0,a_1})$. Here, $M_{b}^{B=i}$ are projective operators such that $\sum_{i\in\{0,1\}}{M_{b}^{B=i}}=\mathbb{I}$. The primary security parameter is the average success probability for $(2\to1)$ QRAC,
\begin{equation}
P_B=\frac{1}{8}\sum_{a_0,a_1,b\in\{0,1\}}{P(B=a_b|a_0,a_1,b)}.
\end{equation}
Parties randomly choose some of the rounds and announce $a_0$ and $a_1$ for those rounds in order to estimate $P_B$. Later the parties perform standard error correction and privacy amplification to obtain perfectly correlated, secure bit strings (see Fig. \ref{21}).

$P_b$ is the only security parameter if we consider the ideal case with perfect detectors i.e. all systems leaving Alice's lab at detected at Bob's end. In the case with losses, the average detection efficiency ($\eta_{avg}$) of Bob's detectors forms the other security parameter. It is important to specify how the communicating parties deal with the rounds in which no particle is detected. Although there are other options, here we choose the simplest one: these rounds are discarded from the statistics. This choice enables the parties to have the estimated average success probability close to the optimal one ( $\frac{2+\sqrt[]{2}}{4}$ in the perfect case for $(2\to1)$ QRAC).
\section{\label{sec:level1} Assumptions and attacks}

We make the following assumptions:
\begin{enumerate}
\item Eve cannot influence the dimension of the system leaving Alice's lab,
\item Eve performs individual attacks,
\item For each bit of the key, Eve's information about it is stored in a bit representing her best guess of this bit,
\item Alice's and Bob's devices are controlled by Eve. She can make detectors work with 100\% efficiency if she chooses to. She also can send information to them  by hidden side channels.  This implies that the states leaving Alice's lab can depend on Eve's choice of measurement. While the measurement basis of Bob's device can depend on both Eve's choice of measurement as well as her outcome,
\item There is no information leakage from the devices. This implies that Eve cannot receive any useful information using hidden side channels.
\item Bob's observed detection efficiency is the same for each of his measurements.
\end{enumerate}

In this part of the paper we study the security of SDIQKD against two distinct classes of attacks,
\begin{enumerate}
\item  Intercept/Resend (without quantum memory),
\item  Delayed Measurement (with a qubit of quantum memory).
\end{enumerate}
\subsection{\label{sec:level2}Intercept/Resend (IR)}
\begin{figure}[htp]
\centering
\resizebox{.95\linewidth}{!}{\includegraphics[]{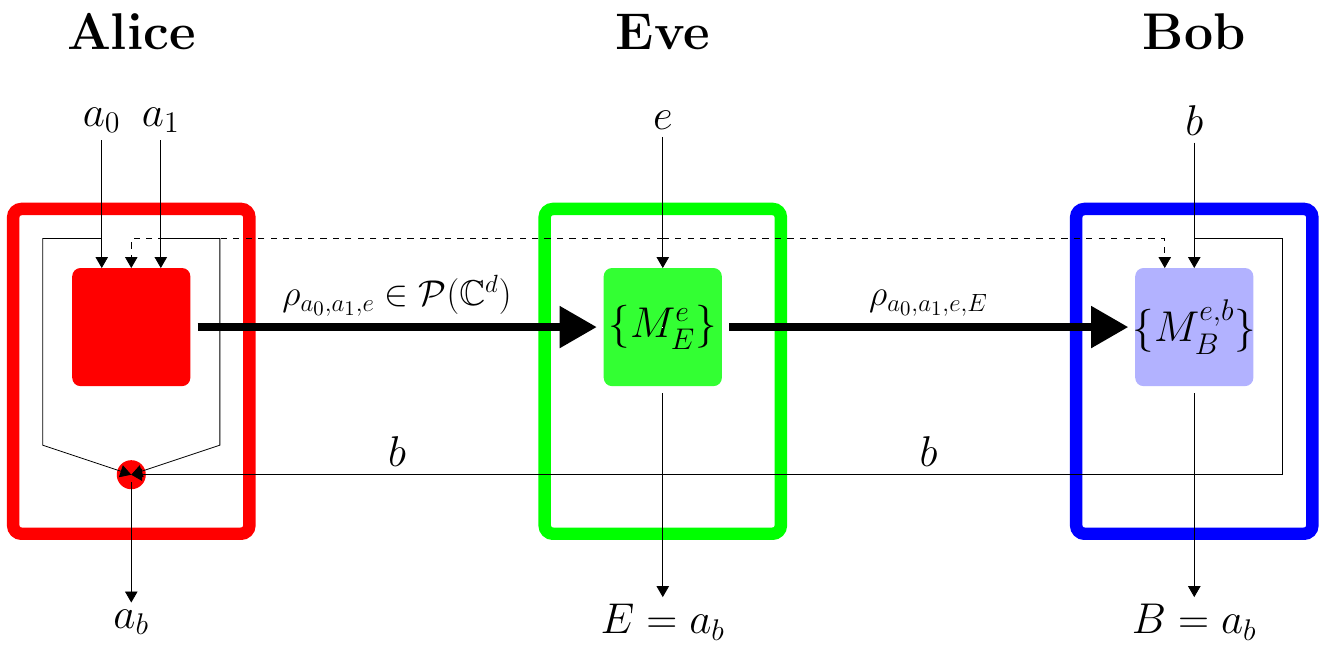}}
\caption{ A snapshot of a successful implementation of the Intercept/Resend (IR) attack on SDIQKD protocol. Here again the thin lines represent classical communication channels while the thick lines represent the quantum communication channel. The dotted lines represent channel for Eve's active control of Alice's and Bob's devices or equivalently channel for distribution of shared randomness. Notice that only the quantum channel leaving Alice's lab has an upper bound on capacity. Finally, as Eve does not have access to quantum memory she has to output her guess of $a_b$ as soon as she intercepts Alice's communicated state. Therefore classical communication carrying Bob's input $b$ is of no use to her except in classical post-processing.}
\label{22}
\end{figure}

 Eve intercepts the signal transmitted from Alice to Bob and measures it in a bases chosen based on her input $e\in \{0,1\}$ (see Fig. \ref{22}) \cite{IR1,IR2}. Eve's input $e$ represents her guess of input of Bob's input $b$ for that particular round. It is crucial for the security of the protocol that $e$ and $b$ are uncorrelated, in other words that there's no information leaking from Bob's lab. However, Eve being able to choose different detection probabilities for rounds when $e=b$ and $e\neq b$ artificially introduces correlations between $e$ and $b$ at the level of post-selected rounds of the experiment.

Eve uses the measurement $M_e^E$ and obtains an output bit $E\in\{0,1\}$.  At this stage we can write Eve's outcome probabilities as,
\begin{equation}
\begin{split}
P(E=i|a_0,a_1,e,b)&=P(E=i|a_0,a_1,e)\\
&=Tr(M_{e}^{E=i}\rho_{a_0,a_1,e}),
\end{split}
\label{Eve'sP}
\end{equation}
where the first equality is because of the fact that Eve gets her outcome $E$ before Bob inputs $b$.
 Here, $M_{e}^{E=i}$  are projection operators such that $\sum_{i\in\{0,1\}}{M_{e}^{E=i}}=\mathbb{I}$.
In turns out that it is optimal for Eve to send the state $\rho_{a_0,a_1,e,E=i}=M_{e}^{E=i}$ to Bob, with probability $P(E=i|a_0,a_1,e)$ as it represents her best knowledge about Alice's input. In fact, it is the most general strategy. Since we assume that the eavesdropper has full control over Bob's measurements, any unitary transformation of the state can be replaced by a corresponding transformation of the measurement bases.

According to assumption 4 Eve, for an observed value of the average detection efficiency of Bob's detectors $\eta_{avg}$  can design in advance all the states $\rho_{a_0,a_1,e}$ and all measurements $M_{e,E,b}^B$ and $M_{e}^E$. However, in any given round of communication she is not aware of values of $a_1$, $a_2$ and $b$ having just $e$ chosen by her.

At Bob's end, $M_{e,E,b}^{B=i}$ are projective operators satisfying $\sum_{i\in\{0,1\}}{M_{e,E,b}^{B=i}}=\mathbb{I}$. Bob's outcome probabilities are given by,
\begin{equation}
P(B=i|a_0,a_1,e,b,E)=Tr(M_{e,E,b}^{B=i}\rho_{a_0,a_1,e,E}).
\end{equation}
As Bob does not know Eve's output, we can obtain the outcome probabilities apparent to Bob, by summing over the values of $E$ which yields,
\begin{eqnarray}
&& P(B=i|a_0,a_1,e,b)= \nonumber \\
&& = \sum_{j\in\{0,1\}}P(E=j|a_0,a_1,e)P(B=i|E=j,a_0,a_1,e,b)\nonumber \\
&& = \sum_{j\in\{0,1\}}Tr(M_{e}^{E=j}\rho_{a_0,a_1,e})Tr(M_{e,j,b}^{B=i}M_{e}^{E=j}).
\end{eqnarray}
Both Eve and Bob are interested in guessing $a_b$, therefore we can write these probabilities using a simplified notation as,
\be
P_{E_b}^{e}=\frac{1}{4}\sum_{a_0,a_1\in\{0,1\}}P(E=a_b|e,a_0,a_1), \label{avgnoteva}\\
P_{B_b}^{e}=\frac{1}{4}\sum_{a_0,a_1\in\{0,1\}}P(B=a_b|e,b,a_0,a_1)\label{avgnotbob}. 
\ee
Next we split Bob's detection efficiency $\eta_{avg}=P(Click)$ into
 \be
 \eta&=&P(Click|e\not=b), \nonumber\\
\eta_{e=b} &=&P(Click|e=b).
 \ee
Here Click signifies occurrence of the event namely Bob's detectors provide an outcome. The other no-click event could occur because of certain malfunction in the set-up (the devices or the channel) or an deliberate attempt at hacking the protocol by a malicious third party. Note that different values of $\eta$ and $\eta_{e=b}$, together with assumption 5, imply that the distribution of $e$ must be uniform.

At this point Eve maximizes $\eta_{e=b}$ making it unity as she wants Bob's device to return the outcomes as often as possible when she managed to guess Bob's input correctly. At the same time she also tries to minimize $\eta$. Only thing limiting her in doing so is the observed detection efficiency which can be easily verified by Bob. Since Eve has no control over Bob's settings, $P(b)=P(b=e)=\frac{1}{2}$. This leads to
\be
\eta_{avg}=\frac{1+\eta}{2}. \label{eta1}
\ee

Here the observed success probabilities for Bob and Eve, post-selected to rounds when Bob's detector registered a particle can be represented as weighted averages over inputs $e$ and $b$,
\be
P_E(\eta)=\frac{1}{2(1+\eta)}\left(P_{E_0}^0+\eta P_{E_1}^0+\eta P_{E_0}^1+P_{E_1}^1\right),\label{PE}
\ee
\be
P_B(\eta)=\frac{1}{2(1+\eta)}\left(P_{B_0}^{0}+\eta P_{B_1}^{0}+\eta P_{B_0}^{1}+P_{B_1}^{1}\right).\label{PB}
\ee

Alice and Bob can establish a secret key if Shannon's mutual information between Alice and Bob is greater than between Alice and Eve ($I(A:B) > I(A:E)$). Assumption 3 makes $E$ a binary observable as it contains her best guess of the $b$th bit of Alice ($a_b$). Then $I(A:E)$ becomes $H(a_b)-H(a_b|E)=1-h(P_E)$, where $h(.)$ is binary Shanon entropy and $P_E$ the probability that $E=a_b$. Similarly $I(A:B)=1-h(P_B)$. The condition $I(A:B) > I(A:E)$ implies $h(P_B)<h(P_E)$. The parties always abort the protocol if the average success probability $P_B$ is lower than the classical maximum winning probability of a $(2\to 1)$ QRAC ($\frac{3}{4}$) as no security can be guaranteed in this case even for perfect detectors \cite{pawlowski2011semi}. Assuming $P_B>\frac{3}{4}$, $P_E>\frac{3}{4}$, in this region the Shannon's entropy function, $h(.)$ is monotonically decreasing, which enables us to simplify $h(P_B)<H(P_E)$ to
\begin{equation}
P_B(\eta)>P_E(\eta). \label{marker}
\end{equation}
Therefore whenever $P_B(\eta)$ is higher than maximal success probability achievable by Eve, $P_E^{max}(\eta)$  protocol is secure. Alice and Bob can rest assured that the protocol is secure if the value of $P_B(\eta)$ is greater than the critical value $P_B^{\mathcal{C}}(\eta)=\frac{1}{2}\max\{P_B(\eta)+P_E(\eta)\}$. Now as both $P_E(\eta)$ and $P_B(\eta)$ are simultaneously maximized, this boils down to finding,  
\be
P_B^{\mathcal{C}}(\eta)=P_E^{max}(\eta)=\max\left\{\frac{P_{E_0}^0+\eta P_{E_1}^0+\eta P_{E_0}^1+P_{E_1}^1}{2(1+\eta)}\right\}.\label{max2}
\ee
with  $P_{E_b}^e=\frac{1}{4}\sum_{a_0,a_1}Tr (\rho_{a_0,a_1,e}M^{E=a_b}_e)$ and the maximization is over all possible measurements of Eve and Bob as well as preparations of Alice. Here we consider two cases:
\begin{itemize}
\item \textit{The general case:} Assumption 4. implies that Eve could have access to shared randomness which allows her to control both the devices during the protocol. We assume that the shared randomness used by Eve in both Alice's and Bob's device is the same as her input $e$. In Alice's device this implies that there are eight possible preparations $\rho_{a_0,a_1,e}$ which depend on Alice's input $a_0,a_1$ as well as Eve's shared random bit $e$. The security condition here is, 
\begin{equation} \label{maxMe}
P_B^{\mathcal{C}}(\eta)=\frac{1}{2}(1+\frac{1}{1+\eta}).
\end{equation}
The details of the physical implementation of the attack and the derivation of the condition can be found in APPENDIX (i),

\item \textit{A minimal characterization of preparation device:} Because of the fact that manipulations in Alice's lab are much more difficult for Eve than just taking control over Bob's lab by hijacking the signal, we start with an assumption that while Eve can choose Alice's preparations, she cannot modify them during the protocol. This assumption is justifiable as the only reasonable strategy to actively control Alice's device is to use shared randomness (or some classical signal which can be modeled using shared randomness) and Alice can use some of her seed to exhaust correlation between her device and Eve's input (or in case of control via classical signal, Alice could easily bar all input signal as her device's only job is to send information). 
This let's us denote Alice's preparations as $\rho_{a_0,a_1}$ as now the state leaving Alice's device is only dependent on her inputs $a_0,a_1$ and not on Eve's input $e$ or shared randomness. Now as Eve wants to maximize her probability of guessing $b$th bit of Alice, it is optimal for her to choose the preparations to be Mutually Unbiased Bases (MUBs). This yields the following security condition, 
\begin{equation} \label{m1}
P_B^{\mathcal{C}}(\eta)=\frac{1}{4}\left(2+\cos \alpha_\eta+\frac{1-\eta}{1+\eta}\sin\alpha_\eta\right).
\end{equation} 
The details of implementation of the attack and the derivation of the security condition can be found in APPENDIX (ii).	
\end{itemize}

\subsection{\label{sec:level2}Delayed Measurement (DM)}
\begin{figure}[htp]
\centering
\resizebox{.95\linewidth}{!}{\includegraphics[]{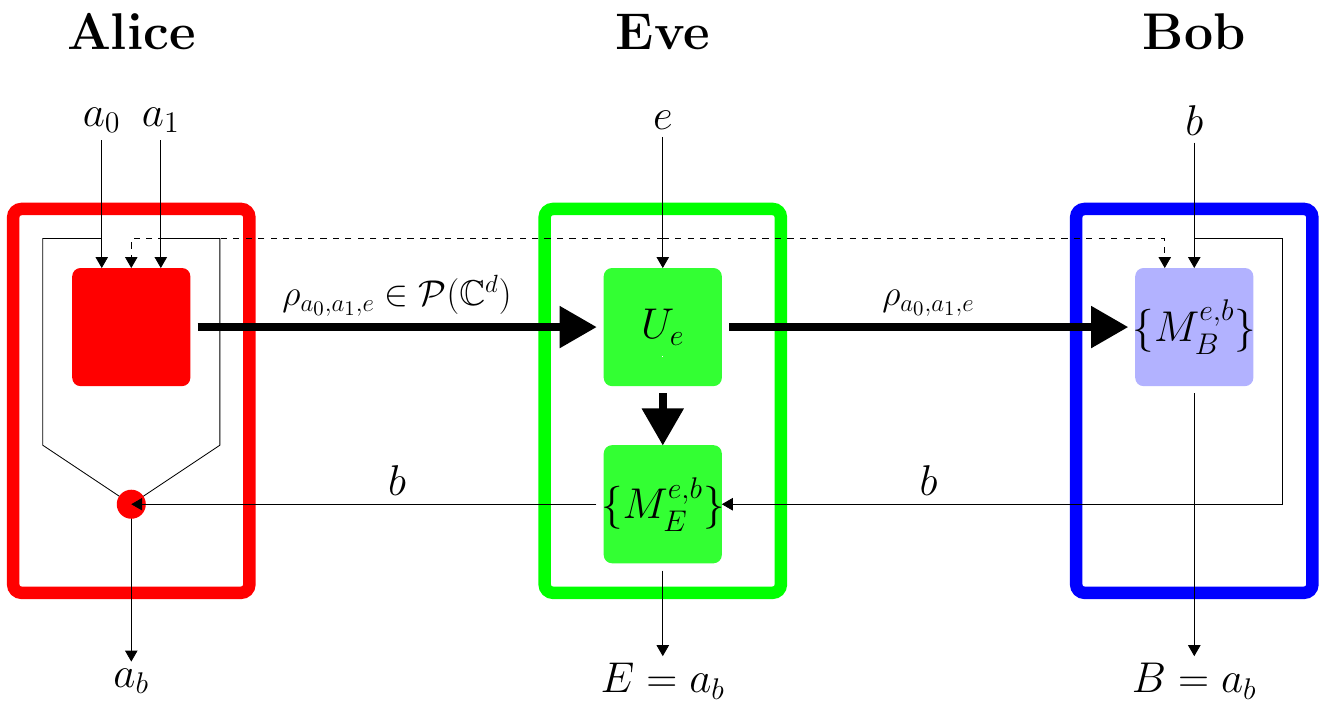}}
\caption{ Successful implementation of Delayed Measurement (DM) attack on SDIQKD protocol. 
Eve's lab now has two devices. The first device intercepts the signal sent from Alice's lab and applies a unitary based on Eve's input bit $e$ on the joint system of the signal and quantum memory. This device then forwards the signal to Bob and quantum memory to Eve's second device. The second device performs a measurement on the quantum memory based on Eve's input $e$, Bob's input $b$ retrieved from classical communication carrying it and yields an output. }
\label{24}
\end{figure}

\begin{figure}
\centering
\resizebox{.95\linewidth}{!}{\includegraphics[]{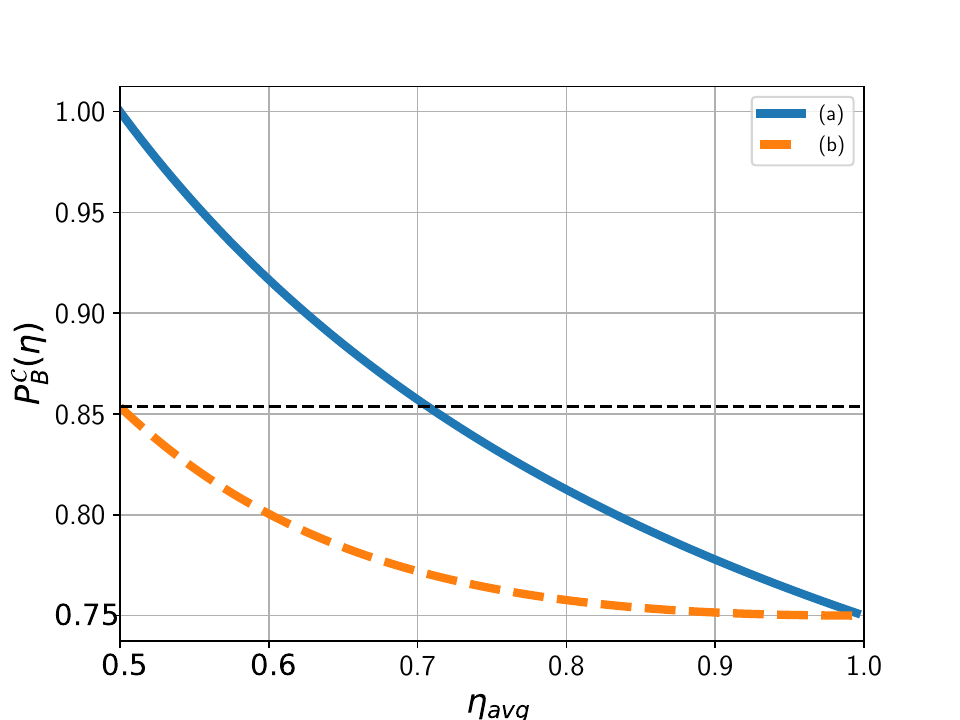}}
\caption{Critical value of the success probability $P_B^{\mathcal{C}}(\eta)$ for SDIQKD based on $(2\to1)$ QRAC vs. observed average detection efficiency $\eta_{avg}$. 
Because we assume that the communicating parties use post-selection and remove all the rounds when Bob did not register a particle from statistics, inefficient detectors do not lower Bob's success probability which can be close to the quantum maximum of $\frac{2+\sqrt{2}}{4}$ (the thin dotted horizontal line) regardless of the value of $\eta_{avg}$. The graphs represent the minimal value of observed success probability $P_B$ required in order to guarantee security against (a) Eve with an unrestricted active control of both Alice's preparation device and Bob's measurement and equipped with (DM) or without (IR) quantum memory and , (b) Eve with no active control of Alice's preparation device and equipped with (DM) or without (IR) quantum memory.. The plot allows one to infer about level of security provided by the devices. This can be done by comparing the observed operational parameters $P_B,\eta_{avg}$ with $P_B^C(\eta)$. If for an observed $\eta_{avg}$, $P_B>P_B^C(\eta)$ then the protocol is secure.
 }
\label{maxpe}
\end{figure}

Let us now consider a more general approach with a more powerful Eve who is equipped with a qubit of memory $\rho_{o}$ (per signal) \cite{DM1,DM2}.  Yet again we deal with two sub-cases
\begin{itemize}
\item \textit{The general case:} Eve after receiving the signal $\rho_{a_0,a_1,e}$ from Alice, without any knowledge about Bob's input $b\in\{0,1\}$ performs unitary transformation $U_e$ (based on her input bit $e$) on both qubits and produces,
\be
\tilde{\rho}_{a_0,a_1,e}=U_e\rho_{a_0,a_1,e}\otimes \rho_{o}U_e^\dagger,
\ee
where $\tilde{\rho}_{a_0,a_1,e}$ is a two qubit state. Eve then forwards the first subsystem to Bob, while holding on to the second one. In this way Eve delays her measurement until Bob publicly announces his setting $b$ (see Fig.\ref{24}).  Bob's projective measurement $M_{b,e}^B$ and Eve's measurement $M_{e,b}^E$ are designed by Eve so as to maximize her success probabilities while keeping Bob's success probabilities to a observed value greater than the threshold classical value. Note that, since Eve now measures after Bob does, Bob's measurement cannot depend on $E$. The joint probabilities can be written as,
\begin{equation}
\begin{split}
P(E=i,B=j|a_0,a_1,e,b)=
\\Tr((M_{e,b}^{E=i}\otimes M_{e,b}^{B=j})\tilde{\rho}_{a_0,a_1,e}).
\end{split}
\end{equation}
Summing over Bob's outcome yields the probability of Eve outcomes,
\begin{equation}
\begin{split}
P(E=i|a_0,a_1,e,b)= \\ Tr((M_{e,b}^{E=i}\otimes \mathbb{I})\tilde{\rho}_{a_0,a_1,e}). 
\end{split}
\end{equation}
In this case we get same final security condition as (\ref{maxMe}) details of which are provided in APPENDIX (iii).
\item \textit{A minimal characterization of preparation device:} The restriction that while Eve can choose the states that leave Alice's lab she cannot alter them during the protocol is still helpful. We find that the optimal states are still the MUBs (\ref{c}) and the security condition we obtain here is the same as (\ref{m1}). 
\end{itemize}
These results were verified using techniques such as the seesaw method based Semi Definite Programming (SDP) \cite{dist1,dist2,dist3} deploying generalized measurements (POVMs) and plotted in Fig. \ref{maxpe}. We conclude that neither quantum memory or generalized measurements help the eavesdropper. 

\section{\label{sec:level1} Modified SDI protocol}

Here we present SDI protocol based on $(3\to1)$ QRAC which is a straightforward generalization of the $(2\to1)$ QRAC and study its security against both (IR and DM) attacks. In a $(3\to1)$ QRAC Alice is given three bits $a_0,a_1,a_2\in {\{0,1\}}$ depending on which she sends the state $\rho_{a_0,a_1,a_2}$, while Bob gets a classical trit, $b\in \{0,1,2\}$ and is required to guess the value of $a_b$. Bob's final output is  $B\in \{0,1\}$ and the success probability is labeled by
$P_B=P(B=a_b)$. The quantum maximum success probability is $\frac{3+\sqrt[]{3}}{6}$ where as the classical maximum remains the same $\frac{3}{4}$. Yet again, the bit $a_b$ forms the raw key bit and after classical post-processing yields the final key. Eve wants to learn $a_b$ in order to establish the same key with Alice as Bob.
\begin{figure}[htp]
\centering
\resizebox{.95\linewidth}{!}{\includegraphics[]{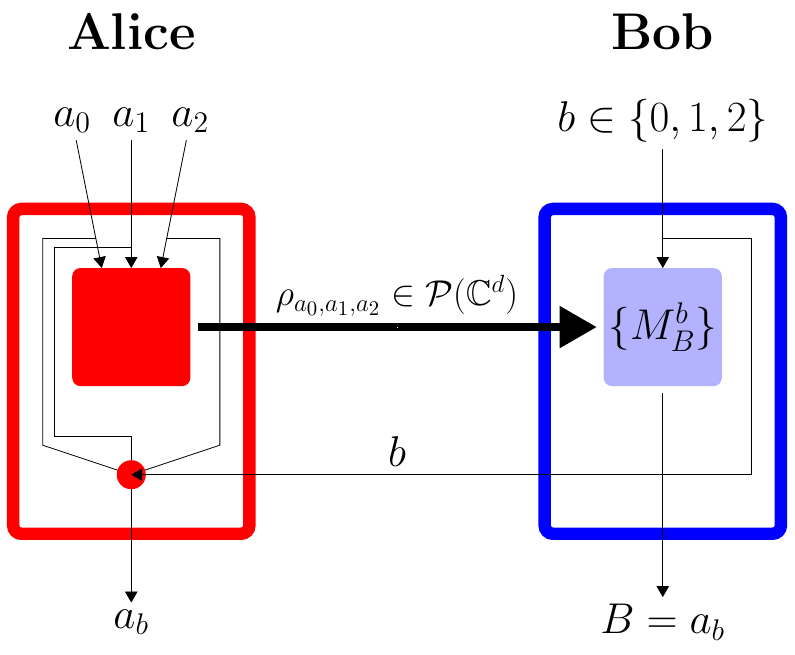}}
\caption{A single round of SDIQKD protocol based $(3\to1)$ QRAC. Alice has three inputs $a_0,a_1,a_2$ instead of two in the $(2\to1)$ QRAC. }
\label{26}
\end{figure}
We keep the structure, the reasoning and the notation (IR and DM) the same as in the SDI protocol based on $(2\to 1$) QRAC. In this case the average detection efficiency of Bob's detector is given by,
\be
\eta_{avg}=\frac{1+2\eta}{3}.
\ee
Owing to the fact that $P(b=e)=P(b)=\frac{1}{3}$. Eve's success probability is given by,
\begin{eqnarray}
&& P_E(\eta)= \frac{1}{3(1+2\eta)}\Big(P_{E_0}^0+\eta(P_{E_1}^0+P_{E_2}^0)+ \nonumber \\
&& P_{E_1}^1+\eta( P_{E_0}^1+P_{E_2}^1)+P_{E_2}^2+\eta( P_{E_1}^2+P_{E_0}^2)\Big)\label{max3}.
\end{eqnarray}
Again we branch into two cases:
\begin{itemize}
\item \textit{The general case:}  Assumption 4. implies that Eve could have access to shared randomness which allows her to control both the devices during the protocol. We assume that the shared randomness used by Eve in both Alice's and Bob's device is the same as her input $e$. In Alice's device this implies that there are eight possible preparations $\rho_{a_0,a_1,a_2,e}$ which depend on Alice's input $a_0,a_1,a_2$ as well as Eve's shared random bit $e$.
Under both IR and DM attacks we have the following security condition,
\begin{equation} \label{3to1max}
P_B^{\mathcal{C}}(\eta)=\frac{1}{2}\bigg( 1 + \frac{1}{1+2\eta}  \bigg).
\end{equation}
In a nutshell, the deviation from \eqref{maxMe} could be attributed to the spread of $e,b\in \{0,1,2\}$. As both $e,b$ are considered to be uniformly random, the chances of them being equal are lowered down to  $P(e=b)\frac{1}{3}$.  
The details of the implementation and a brief proof sketch is provided in APPENDIX (iv).

\item \textit{A minimal characterization of preparation device} As manipulations in Alice's lab are much more difficult for Eve than just taking control over Bob's lab by hijacking the signal, we start with an assumption that while Eve can choose Alice's preparations, she cannot modify them during the protocol. 
This let's us denote Alice's preparations as $\rho_{a_0,a_1,a_2}$ as now the state leaving Alice's device is only dependent on her inputs $a_0,a_1,a_2$ and not on Eve's input $e$ or shared randomness. This yields the following security condition for both IR and DM attacks is as follows, 
\be
P_B^{\mathcal{C}}(\eta)=
\begin{cases}
\phi(\eta), & \text{for}\ \eta \in [0,\frac{3\sqrt{2}-4}{2}]\\
\frac{3}{4}, & \eta \in (\frac{3\sqrt{2}-4}{2},1] \label{max3to1res}
\end{cases}
\ee
where
\be \nonumber
\phi(\eta)&=&\frac{1}{8}\Big(4+(1+\cos \alpha_\eta)\cos\beta_\eta+ \nonumber\\  &+&\frac{2(1-\eta)}{1+2\eta}\sin\alpha_\eta\sin\beta_\eta\Big), \label{e4} \nonumber
\ee
\begin{equation} \nonumber
\alpha_\eta=\arccos\left(\frac{1}{N(\eta)^2-1}\right),
\end{equation}
\begin{equation} \nonumber
\beta_\eta=\arctan\left(\frac{\tan(\alpha_\eta)}{N(\eta)}\right),
\end{equation}
\begin{equation}
N(\eta)=\frac{2(1-\eta)}{1+2\eta}.
\end{equation}
Unlike the previous case, if Eve wants to maximize her probability of guessing $b$th bit of Alice, it is optimal for her to choose the preparations that are not MUBs but converge to MUBs under a specific efficiency condition. 
The details are provided in APPENDIX (v).
\end{itemize}
These results were verified using techniques such as the seesaw method based Semi Definite Programming (SDP) deploying generalized measurements (POVMs) and plotted in Fig. \ref{maxpe3}.
\begin{figure}[!]
\centering
\resizebox{.95\linewidth}{!}{\includegraphics[]{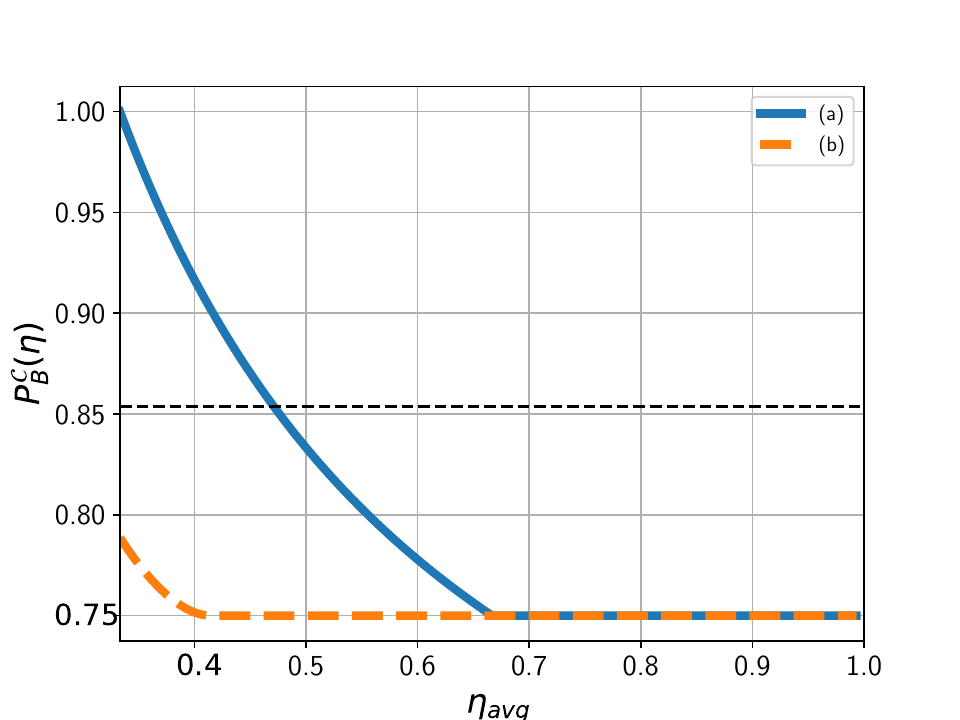}}
\caption{Critical value of the success probability $P_B^{\mathcal{C}}(\eta)$ for SDIQKD based on $(3\to1)$ QRAC vs. $\eta_{avg}$. The thin dotted horizontal line represents the quantum maximum winning probability $\frac{3+\sqrt[]{3}}{6}$.
The graphs represent the minimal value of observed success probability $P_B$ required in order to guarantee security against (a) Eve with an unrestricted active control of both Alice's preparation device and Bob's measurement and equipped with (DM) or without (IR) quantum memory and , (b) Eve with no active control of Alice's preparation device and equipped with (DM) or without (IR) quantum memory. The plot allows one to infer about level of security provided by the devices. This can be done by comparing the observed operational parameters $P_B,\eta_{avg}$ with $P_B^C(\eta)$. If for an observed $\eta_{avg}$, $P_B>P_B^C(\eta)$ then the protocol is secure. Notice the increased tolerance (possibility of a secure protocol) at lower values of detection efficiency ($\eta_{avg}$) as compared to the security offered by the protocol based on $(2\to 1)$ QRAC Fig. 4. }
\label{maxpe3}
\end{figure}

\section{\label{sec:level1}Conclusions.}

In this paper we analyzed individual {\it quantum hacking} attacks on SDIQKD protocols based on QRACs where eavesdropper can not only design but actively control all devices during the protocol. Looking at these types of attacks was motivated by their recent experimental realizations. We study security against two types of quantum eavesdroppers (with and without access to quantum memory) and for two distinct levels of characterizations of the devices (with and without a minimal characterization of the preparation device). We found that access to small quantum memory (a qubit) does not help the eavesdropper to attack the SDIQKD protocol and conjecture that the same holds for other protocols and unlimited memory.  As QKD in general is gaining immense popularity \cite{QKD1,QKD2,QKD3,QKD5} and entanglement based QKD remains commercially nonviable, devices employing prepare and measure QKD schemes seems to be natural way forward. Our analysis other than being robust, deals with worst case scenarios and provides for the everyday naive user a hassle-free way to infer about the security his devices offer. In particular Fig. 4. and Fig. 6. provide a straightforward way to ensure security, namely, a user can crosscheck the operational security parameters $P_B,\eta_{avg}$ against ${P^C_B(\eta)}$, and if he finds the $P_B>P_B^C(\eta)$, he can rest-assured without going into further details. Our double-layered results enable such crosschecking for two layers of device specification and may be used according to varying degrees of trust in the provider.    

We provide condition for establishing a secure key for SDIQKD based on $(2\to1)$ QRAC and $(3\to1)$ QRAC against Eve who has full active control of their devices. Using SDIQKD based on $(3\to1)$ QRAC lowers the key-rate but the security requirements are significantly lowered. Further, a minimal characterization for the preparation device is provided which lowers the critical detection efficiency all the way down to $50\%$ for $(2\to1)$ QRAC and to $41.2\%$ for $(3\to1)$ QRAC. We have listed the critical detection efficiencies for the various cases considered in Table \ref{Table1}.
It is known that $(2\to1)$ RAC and $(3\to1)$ RAC can also be implemented using entanglement and classical communication, often called the $(2\to1)$ Entanglement Assisted RAC (EARAC). These implementations can also be seamlessly used for QKD using the aforementioned method. We conjecture based on numerical evidence that the results derived in this work still hold for $(2\to1)$ and $(3\to1)$ EARAC.

\begin{table}[]
\centering
\begin{tabular}{c|c|c}
$\eta_{avg}^{critical}$ & (a)  & (b)  
        \\ \hline
$(2\to1)$ QRAC          & 0.71 & $\frac{1}{2}$ \\ \hline
$(3\to1)$ QRAC          & 0.58 & $\frac{1}{3}$ 
\end{tabular}
\caption{Critical average detection efficiency $\eta_{avg}^{critical}$ for (a) Eve without quantum memory (IR) or equipped with quantum memory (DM) and unrestricted active control of both Alice's preparation device and Bob's measurement device, (b) Eve without quantum memory (IR) or equipped with quantum memory (DM) and no active control of Alice's preparation device. }
\label{Table1}
\end{table}

We would like to remark that the critical detection efficiency is the efficiency of the whole process taking into the account not only the losses in the device of the receiver but also in the transmission. Therefore, in practice, they will increase with the distance between the parties and the critical detection efficiency of a protocol puts a bound on how far apart the communicating parties can be. For the standard device independent QKD this distance is just a couple of kilometers \cite{dist}. Using SDI protocols described here it can be significantly extended. Our results suggest a connection between security and MUB based encoding decoding schemes, which deserves further exploration. While this work studied security of SDIQKD protocols with constrained capacity (dimension) of the communication channel, security based on other SDI constraints have also shown potential, for instance the oblivious constraint as introduced in \cite{OC}. 

\section{acknowledgments}

This project was supported by a grant, FirstTEAM
(Grant No. FirstTEAM/2016-1/5) from FNP, ERC
AdG QOLAPS.

\clearpage
\appendix
\section*{Appendix}
\subsection*{(i)}
W.l.o.g. we can consider Eve's projective measurements to be, 
\begin{eqnarray}
&& |M_e^{E=0}\rangle=\cos\frac{\alpha_e}{2}|0\rangle+\sin\frac{\alpha_e}{2}|1\rangle \nonumber \\
&& |M_e^{E=1}\rangle = |M_e^{E=0}\rangle^\perp \label{a4},
\end{eqnarray}
and $M_e^{E=0}=|M_e^{E=0}\rangle \langle M_e^{E=0} |$,$M_e^{E=0}=|M_e^{E=1}\rangle \langle M_e^{E=1}|$. Now one can rewrite (\ref{max2}) as,
\be \label{coolway!}
P_E^{max}(\eta) = \frac{1}{2}\max\left\{P_E^{a_0=a_1} + P_E^{a_0\neq a_1}(\eta) \right\},
\ee
where
\begin{equation} \label{subset1}
\begin{split}
P_E^{a_0 = a_1}=&
\frac{1}{4} Tr\bigg(\rho_{000}M^0_0 + \rho_{001}M^0_1 + \\
& \rho_{110}M^1_0 + \rho_{111}M^1_1\bigg), \\ 
\end{split}
\end{equation}
\begin{equation} \label{subset2}
\begin{split}
P_E^{a_0 \neq a_1}(\eta)= & \frac{1}{4} Tr\bigg(\rho_{010}\frac{M^0_0+\eta M^1_0}{1+\eta} + \\ & \rho_{011}\frac{M^1_1+\eta M^0_1}{1+\eta} + \rho_{100}\frac{M^1_0+\eta M^0_0}{1+\eta} + \\ & \rho_{101}\frac{M^1_1+\eta M^0_1}{1+\eta}\bigg.).
\end{split}
\end{equation}
Notice that (\ref{subset1}) and (\ref{subset2}) divide Alice's preparations into two mutually exclusive subsets. Alice's states that maximize $P_E^{a_0=a_1}$, $\rho_{0,0,0},\rho_{0,0,1},\rho_{1,1,0},\rho_{1,1,1}$ remain the same irrespective of whether Eve was able to correctly guess Bob's input ($e=b$) or not ($e\neq b$) simply because both of Alice's input are the same. This allows Eve to set these states equivalent to the projectors $M_0^0,M_1^0,M_0^1,M_1^1,$ respectively.   Which in-turn allows one to re-write (\ref{coolway!}) as,
\begin{equation}
\begin{split} \label{labelme}
P_E^{max}(\eta) = \frac{1}{2}(1+\max\{P_E^{a_0\neq a_1}(\eta)).
\end{split}
\end{equation}
Now in-order to find the maximum value of (\ref{subset2}) consider one of the terms involved,
\begin{equation}
\begin{split}
Tr\bigg(\rho_{010}\frac{M_0^0 + \eta M_0^1}{1+\eta}\bigg) = \frac{1}{1+\eta} - \frac{1-\eta}{1+\eta}Tr\bigg(\rho_{010}M_0^1\bigg),
\end{split}
\end{equation}
where the equality stems from the fact that $M_0^0=\mathbb{I}-M_0^1$.
The maximum for this term is reached by setting $\rho_{010}=M_0^0$ which yields the final security condition (\ref{maxMe}).
\subsection*{(ii)}
W.l.o.g she fixes Alice's preparations to be MUBs,
\begin{eqnarray}
&& \rho_{00}=|0\rangle\langle0|, \nonumber \\
&& \rho_{01}=\frac{1}{2}(|0\rangle+|1\rangle)(\langle0|+\langle1|), \nonumber \\
&& \rho_{10}=\frac{1}{2}(|0\rangle-|1\rangle)(\langle0|-\langle1|), \nonumber \\
&& \rho_{11}=|1\rangle\langle 1|, \label{c}
\end{eqnarray}
which is an optimal set of states for the standard ($2\to1$) QRAC. Next w.l.o.g. we characterize Eve's projective measurement using vectors from the same plane as the states in (\ref{c}),
\begin{eqnarray}
&& |M_e^{E=0}\rangle=\cos\frac{\alpha_e}{2}|0\rangle+\sin\frac{\alpha_e}{2}|1\rangle, \nonumber \\
&& |M_e^{E=1}\rangle = |M_e^{E=0}\rangle^\perp \label{a4},
\end{eqnarray}
and $M_e^{E}=|M_e^E\rangle\langle M_e^E|$. This allows us to partition (\ref{max2}) into two parts based on different values of $e$. These parts are independent and, due to symmetry, equal. Therefore we may re-write $P_E(\eta)$ as ,
\be
P_E(\eta)=\max\left\{\frac{P_{E_0}^0+\eta P_{E_1}^0}{1+\eta}\right\},
\ee 
which is,
\begin{equation}
\begin{split}
P_E(\eta)= &
\frac{1}{8(1+\eta)}Tr\bigg(M_{e=0}^{E=0}(\rho_{00}+\rho_{01})+ \\ & M_{e=0}^{E=1}(\rho_{10}+\rho_{11})\bigg)
+ \\& \frac{\eta}{8(1+\eta)} Tr\bigg(M_{e=0}^{E=0}(\rho_{00}+\rho_{10})+ \\ & M_{e=0}^{E=1}(\rho_{01}+\rho_{11})\bigg).
\end{split}
\end{equation}
After plugging in (\ref{c}),(\ref{a4}), this yields,
\begin{equation}
\begin{split}
P_E^{max}(\eta)& =\max_{\alpha_0}(P_E(\eta)) \\ & = \frac{1}{4}\left(2+\cos \alpha_0+\frac{1-\eta}{1+\eta}\sin\alpha_0\right).
\end{split}
\end{equation}
This expression is maximized for $\alpha_\eta=\arctan\left(\frac{1-\eta}{1+\eta}\right)$. Hence, we obtain the security condition (\ref{m1}).
\subsection*{(iii)}
Here the success probability for Eve is,
\begin{equation}
P_E(\eta)= \frac{1}{8(1+\eta)}Tr \sum_{a_0a_1eb} \eta^{1-\delta_{eb}}P(E=a_b|a_0,a_1,e,b),
\end{equation}
which can in-turn be expressed as,
\begin{equation}
\begin{split}
P_E(\eta)= \frac{1}{8} Tr\bigg[ & \sum_{i,j} \tilde{\rho}_{i,i,j} \frac{M_{i,i}^{E=j} \otimes \mathbb{I} + \eta M_{i,1-i}^{E=j} \otimes \mathbb{I} }{1 + \eta} + \\ & \tilde{\rho}_{i,1-i,j} \frac{M_{i,i}^{E=1-\delta_{i,j}} \otimes \mathbb{I} + \eta M_{i,1-i}^{E=\delta_{i,j}} \otimes \mathbb{I} }{1 + \eta} \bigg].
\end{split}
\end{equation}
Notice that this expression constitutes four independent elements  for specific value of the pair $(i,j)$. In order to find $P_E(\eta)^{max}$ we need only find the maximizing condition for one term. Let's consider a particular pair $(i,j)$, then the expression for $P_E(\eta)^{max}$ simplifies to,
\begin{equation}
\begin{split}
P_E^{max}(\eta)= & \frac{1}{2} \max \bigg\{ Tr\bigg(  \tilde{\rho}_{i,i,j} \frac{M_{i,i}^{E=j} \otimes \mathbb{I} + \eta M_{i,1-i}^{E=j} \otimes \mathbb{I} }{1 + \eta} + \\ & \tilde{\rho}_{i,1-i,j} \frac{M_{i,i}^{E=1-\delta_{i,j}} \otimes \mathbb{I} + \eta M_{i,1-i}^{E=\delta_{i,j}} \otimes \mathbb{I} }{1 + \eta} \bigg) \bigg\}.
\end{split}
\end{equation}
This equation is maximized when $M_{i,i}^{E=j}=M_{i,1-i}^{E=j}$ or $M_{i,i}^{E=1-\delta_{i,j}}=M_{i,1-i}^{E=\delta_{i,j}}$ yielding the same security condition as (\ref{maxMe}).
\subsection*{(iv)}
In the case when Eve does not have access to quantum memory (IR) we can re-write (\ref{max3}) in a convenient way as,
\begin{equation}
P_E^{max}(\eta)=\frac{1}{4}\max \bigg\{P_E^{a_0=a_1=a_2} + P_E^{NOT\{a_0=a_1=a_2\}} \bigg\},
\end{equation}
where $P_E^{a_0=a_1=a_2}$ is Eve's success probability for the case when all three of the input bits of Alice are equal and $P_E^{NOT\{a_0=a_1=a_2\} }$ is Eve's success probability for the case when the three inputs of Alice are not equal. As Alice's states and Bob's measurement that maximize $P_E^{a_0=a_1=a_2}$ remain the same irrespective of the fact whether Eve was able to guess Bob's input correctly or not we can further re-write this as,
\begin{equation}
P_E^{max}(\eta)=\frac{1}{4}\bigg( 1 + \max \bigg\{P_E^{NOT\{a_0=a_1=a_2\}} \bigg\} \bigg).
\end{equation}
Following exactly the same steps as above this yields the security condition (\ref{3to1max}).
\subsection*{(v)}
We find that the optimal states are,
\begin{eqnarray}
&& |000\rangle=|0\rangle,\nonumber\\
&& |001\rangle=\cos \frac{\alpha}{2}|0\rangle+\sin\frac{\alpha}{2}|1\rangle,\nonumber\\
&& |010\rangle=\cos \frac{\alpha}{2}|0\rangle+e^{i\beta}\sin\frac{\alpha}{2}|1\rangle,\nonumber\\
&& |100\rangle=\cos \frac{\alpha}{2}|0\rangle+e^{-i\beta}\sin\frac{\alpha}{2}|1\rangle,\nonumber\\
&& |111\rangle=|000\rangle^\perp ,\nonumber \\
&& |110\rangle=|001\rangle^\perp ,\nonumber \\
&& |101\rangle=|010\rangle^\perp ,\nonumber \\
&& |011\rangle=|100\rangle^\perp,\label{a11}
\end{eqnarray}
where $\alpha$ and $\beta$ are parameters controlled by Eve.  In this case optimal encoding for standard $(3\to1)$ QRAC is reproduced for $\alpha=\arccos\frac{1}{3}$ and $\beta=2\pi/3$.
For Eve's measurements ($ M_e^{E=a_b}=|M_e^{E=a_b}\rangle\langle M_e^{E=a_b}|$) we use the following parametrization
\begin{eqnarray}
&& |M_e^{E=0}\rangle = \cos\frac{\alpha_e}{2}|0\rangle+e^{i\beta_e}\sin\frac{\alpha_e}{2}|1\rangle \nonumber \\
&& |M_e^{E=1}\rangle = |M_e^{E=0}\rangle^\perp \label{a2},
\end{eqnarray} 
A straightforward maximization yields the security condition (\ref{max3to1res}).

\end{document}